\documentclass[aip,jcp,reprint,superscriptaddress]{revtex4-2}
\usepackage{graphicx}
\usepackage{appendix}
\usepackage{xcolor}
\usepackage{mhchem}

\begin{document}
\title{Improving the reliability of machine learned potentials for modeling inhomogenous liquids}
\author{Kamron Fazel}\thanks{Equal contribution}
\author{Nima Karimitari}\thanks{Equal contribution}
\author{Tanooj Shah}
\author{Christopher Sutton}\email{cs113@mailbox.sc.edu}
\author{Ravishankar Sundararaman}\email{sundar@rpi.edu}

\begin{abstract}
The atomic-scale response of inhomogeneous fluids at interfaces and surrounding solute particles plays a critical role in governing chemical, electrochemical and biological processes at such interfaces.
Classical molecular dynamics simulations have been applied extensively to simulate the response of inhomogeneous fluids directly, and as inputs to classical density functional theory, but are limited by the accuracy of the underlying empirical force fields.
Here, we deploy neural network potentials (NNPs) trained to \emph{ab initio} simulations to accurately predict the inhomogeneous response of two widely different fluids: liquid water and molten NaCl.
Although NNPs can be readily trained to model complex bulk systems across a range of state points, in order to appropriately model a fluid's response at an interface, inhomogeneous configurations must be included in the training data.
We establish protocols based on molecular dynamics simulations in external atomic potentials in order to sufficiently sample the correct configurations of inhomogeneous fluids. 
We show that NNPs trained to inhomogeneous fluid configurations can predict several properties such as the density response, surface tension and size-dependent cavitation free energies in water and molten NaCl corresponding to \emph{ab initio} interactions more accurately than empirical force fields. 
This work therefore provides a first demonstration and framework for extracting the response of inhomogeneous fluids from first principles  for classical density-functional treatment of fluids free from empirical potentials.
\end{abstract}

\maketitle

\section{Introduction}

Solvation critically determines the energetics, dynamics and chemistry of molecules in solution and at solid-liquid interfaces.\cite{koperIntroductionComputationalElectrochemistry2022}
Computational approaches to predict solvation vary widely in computational cost and accuracy from explicit molecular dynamics (MD) simulations, including classical and \emph{ab initio} MD (AIMD), to implicit solvation models that reduce the statistically-averaged response of the liquid to that of a continuum dielectric.\cite{schwarzElectrochemicalInterfaceFirstprinciples2020, ringeImplicitSolvationMethods2022}
Improving the accuracy of the treatment of solvation in first-principles calculations requires treatment of the atomic-scale response of fluids, ranging from nonlocal extensions of implicit models,\cite{sundararamanWeighteddensityFunctionalsCavity2014, sundararamanSpicingContinuumSolvation2015} to self-consistent fluid density models including integral-equation approaches, such as 3D Reduced Interaction-Site Model (3D-RISM),\cite{kovalenkoSelfconsistentDescriptionMetal1999} and classical density functional theory (DFT).\cite{wuDensityFunctionalTheoryComplex2007}

All solvation approaches require, by definition, an accurate treatment of the inhomogeneous response of fluids.
Classical MD simulations using empirical potentials have been extensively used to investigate the inhomogeneous response of liquids, including the structure, surface interactions, and surface tension of liquid-vapor interfaces.\cite{yuetMolecularDynamicsSimulation2010, zengMechanisticInsightWater2023, liuLigandEffectSwitching2023, huangScalingHydrophobicSolvation2001}
Particularly relevant for solvation are the structure and free energy involved with forming cavities of different sizes in liquids, which exhibit qualitatively different behavior for small cavities, and approach the behavior of liquid-vapor interfaces only for large cavities.\cite{huangScalingHydrophobicSolvation2001, sundararamanRecipeFreeenergyFunctionals2014}

Cavitation simulations have played a vital role in the development of classical DFT approximations for inhomogeneous fluids,\cite{sundararamanEfficientClassicalDensityfunctional2014, sundararamanRecipeFreeenergyFunctionals2014} which can predict the equilibrium density and energy of fluids in complex environments at a fraction of the computational cost of MD by eliminating the need for statistical sampling.\cite{wuDensityFunctionalTheoryComplex2007}
Previous works \cite{sprousterMolecularStructurePhase2022} have also computed the solvation-relevant properties in molten salt systems.
However, such approaches remain largely dependent on empirical potentials, potentially limiting the accuracy and applicability to a narrow range of liquids and electrolytes.

Recent developments in machine-learned interatomic potentials such as neural-network potentials (NNPs), trained to AIMD simulations, facilitate classical MD-scale simulations free of empirical potentials, with the accuracy of the underlying \emph{ab initio} data.\cite{behlerPerspectiveMachineLearning2016, wangDeePMDkitDeepLearning2018,kocerNeuralNetworkPotentials2022}
Most applications of NNPs to liquids focus on the structure and dynamics of bulk homogeneous liquids, and examine properties like the radial distribution functions (RDFs).\cite{leeComparativeStudiesStructural2021, liangTheoreticalPredictionLocal2021, chahalTransferableDeepLearning2022} Among the list of all NNP methods, deep potential molecular dynamics (DPMD) \cite{wangDeePMDkitDeepLearning2018} is proven to accurately predict all the known phases of water \cite{zhangPhaseDiagramDeep2021}, and inhomogeneous liquids with explicit interfaces, e.g., proton transfer at solid-liquid interface of \ce{TiO2}-water.\cite{andradeFreeEnergyProton2020} These studies require treating solute atoms explicitly and depend heavily on the specific solute atoms. Therefore, they are incapable exploring the general inhomogeneous response of liquids, independent of the type of the solute as needed for classical DFT development.

In this work, we develop a generalized protocol to train, test and apply NNPs to quantify the general, first-principles inhomogeneous response of liquids that is independent of empirical potentials and specific solutes. Applications of NNPs for liquid-vapor interfaces have previously shown the need to include \emph{ab initio} vapor-liquid interface configurations in the training data.\cite{niblettLearningIntermolecularForces2021} To generalize the NNP to arbitrary inhomogeneous configurations, we apply external atomic potentials with varying spatial profiles and strengths to each atomic species in the MD simulations to systematically map out the inhomogeneous response.
We show that correspondingly inhomogeneous AIMD simulations are necessary in the training data in order to obtain reliable inhomogeneous MD simulations with the resulting NNPs.
Using such inhomogeneous NNPs trained to external-potential data, which we denote as `NNP-ext', we predict the density response in external potentials, liquid-vapor surface tension and size-dependent cavitation free energies for liquid water and molten NaCl. This is the first time, to our knowledge, that NNPs have been applied to calculations of cavitation free energies. 
Comparison between the first-principles NNP-ext predictions with those from traditional classical MD predictions highlights the need for extensive NNP MD simulations of inhomogeneous liquids to facilitate the development of accurate first-principles solvation models.

\section{Method overview}

\begin{figure}
\includegraphics[width=\columnwidth]{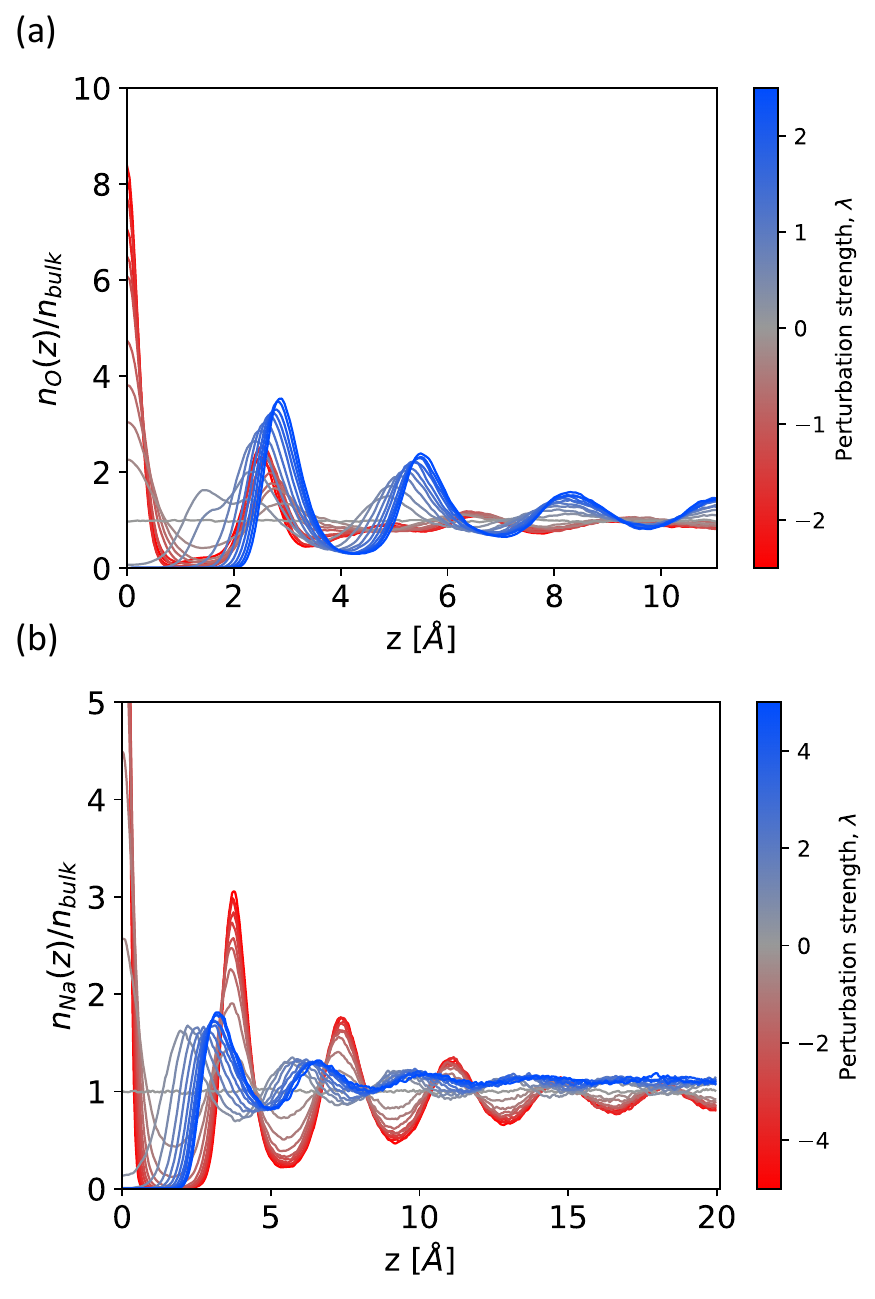}
\caption{Density profiles of (a) water at 298 K and (b) molten NaCl at 1300 K in the NVT ensemble under external potentials for repulsive and attractive planar Gaussian potentials with width 1~\AA, centered at $z=0$, applied to oxygen and sodium atoms, respectively. 
The repulsive potential repels oxygen / sodium from the origin, forming a cavity at strong enough potentials, followed by an oscillating shell structure at larger distances.
The attractive potential creates an increased localized density at $z=0$ with an opposite shell structure a larger distances away from $z=0$.}
\label{fig:densityClassical}
\end{figure}

To generate training data for the NNPs, MD simulations are performed with arbitrary external potential profiles (e.g., planar, Gaussian) applied to each separate atom type to systematically sample the resulting inhomogeneous response. 
For example, Figure~\ref{fig:densityClassical} shows the equilibrium density profiles of water and molten NaCl treated using classical interatomic potentials in planar repulsive and attractive Gaussian potentials with varying strength.
Such choices for the analytical form of the external potentials allows for mapping out the response of liquids at interfaces varying continuously from (for example) hydrophobic to hydrophilic, as the potential varies from repulsive to attractive.
Note that such responses include non-perturbative inhomogeneities of the liquid and contain much more information than RDFs, which typically only provide information about the bulk fluid. 
The inclusion of strong inhomogeneities is critical for accurate solvation, especially at electrochemical interfaces where interface charge can lead to strong potentials on the fluid atoms.

To test our protocol on a range of fluid types we examine liquid water, a molecular fluid of general scientific interest, and molten NaCl, a simple ionic liquid. 
For each system we compare simulations using three different kinds of force fields - standard classical pairwise force fields, NNPs trained to bulk liquid AIMD alone, and NNPs trained to additional data with external-potential AIMD simulations, which we refer to as NNP-ext.
See Appendix~\ref{sec:ClassicalPot} for details on the classical force fields and Appendix~\ref{sec:NNP} for details on the AIMD training sets and training protocols for the NNPs.

Through this work, we implemented new open-source tools for the inclusion of combinations of spherical, cylindrical or planar Gaussian external potentials to classical MD and AIMD simulations. 
We created the open-source framework MDext\cite{sundararamanMolecularDynamicsExternal2023} by extending the Python interface to LAMMPS.\cite{plimptonFastParallelAlgorithms1995} 
This code facilitates collecting density profiles in the same geometry as the external potential, and extracting the quantities required for computing cavitation free energies reported in this work. 
Finally, for the inhomogeneous AIMD simulations to augment the NNP training sets, we implemented the capability to apply the same classes of external potentials in the open-source plane-wave DFT software, JDFTx.\cite{sundararamanJDFTxSoftwareJoint2017}

\section{Results}

\subsection{Force Field Uncertainty Analysis}

\begin{figure}
\includegraphics[width=\columnwidth]{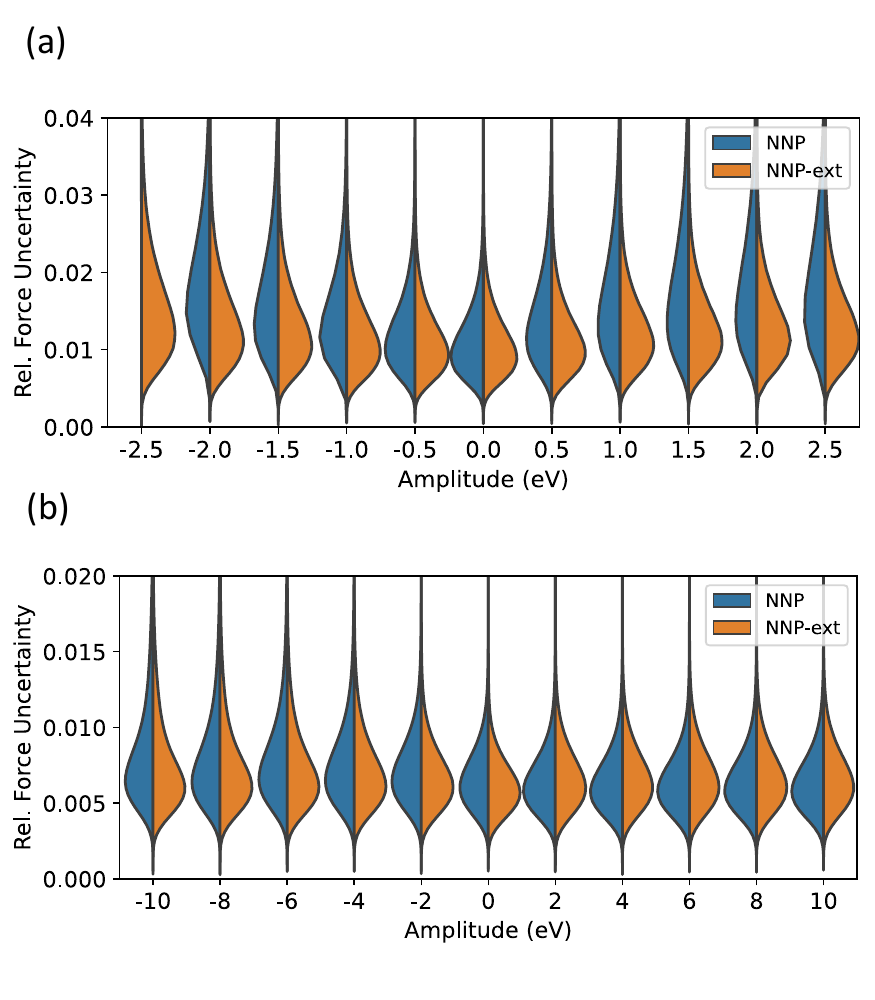}
\caption{
The relative force uncertainty distribution of all atoms under different magnitudes of planar external potentials for (a) water at 298 K and 1 atm and (b) molten salt at 1300 K and 1 atm. For water, the NNP distribution shifts to higher uncertainties compared to NNP-ext, eventually failing at -2.5 eV. However, the opposite happens for molten salt where the distribution of NNP and NNP-ext remains identical at any external potentials.}
\label{fig:unc}
\end{figure}

We start by directly examining the inhomogeneous response of water and molten NaCl in external potentials of varying strengths and compare errors between NNPs and NNP-exts. 
For water, the NNP MD simulations were performed at $T=298 K$ in an NVT ensemble initialized at an equilibrium volume corresponding to $P=1$ bar, with the amplitude of external planar Gaussian potentials ranging from $-2.5$ eV to $+2.5$ eV. 
In all these cases, the systems equilibrated within 50 ps in the MD simulation, after which data was collected for another 100 ps. 

To assess uncertainties between the NNPs and NNP-exts, we train a committee of four potentials with different random seeding for the neural network weights.
The relative error in force is assessed with a root mean square error type method within DeePMDs inference in LAMMPS, as was established in  Ref \citenum{zengDeePMDkitV2Software2023}.
Higher error indicates the NNP has not adequately learned a particular environment.

We calculate the relative force uncertainty per atom in every 50 fs of committee MD simulations. Figure~\ref{fig:unc}a shows the distribution of force uncertainties for all atoms at varying amplitudes of the external potential. 
At zero external potential, that is, in the bulk fluid, both NNP and NNP-ext have identical distribution with a mean value of about 0.01 (or 1 \% relative force error). 
However, as the magnitude of external potential increases and system becomes more inhomogeneous, the NNP have higher uncertainties with shallower distribution compared to NNP-ext. For example at the amplitude of +2.0 eV, the 95 percentile of NNP is at 0.036 about 50 \% larger than the corresponding value for NNP-ext at 0.024. 
The same trend is observed for the attractive external potential, with the exception that for -2.5 eV and stronger attractive potentials, the MD simulation is no longer stable and fails in less than 1 ps. 
The NNP-ext, in contrast, remains stable even under extreme attractive potentials of amplitudes up to -15 eV.

We perform the same uncertainty analysis for molten NaCl in Fig~\ref{fig:unc}b. 
The MD simulations are carried out at $T=1300 K$ in an NVT ensemble initialized at $P=1$ bar, with external planar Gaussian potentials ranging from $-5$ eV to $+5$ eV (larger amplitude than the water case corresponds to the increased thermal energy). 
In contrast to water, molten NaCl  exhibits almost identical uncertainty distributions for both NNP and NNP-ext, with differences of less than 0.001 in the relative force uncertainty. 
Additionally, both NNP and NNP-ext remain stable for extreme attractive external potentials of amplitudes up to -20 eV.

This indicates that the molten NaCl is an easier case for the NNP to capture the inhomogeneous configuration energies, likely because the strongly ionic interaction is simpler to learn with NNPs.
In contrast, water is a much more challenging system with intramolecular covalent bonds, hydrogen bonds and longer range interactions, all of which change from bulk to inhomogeneous systems, necessitating explicit inhomogeneous AIMD to even achieve a stable MD trajectory.

\subsection{Surface tension}

\begin{figure}[ht!]
\includegraphics[width=\columnwidth]{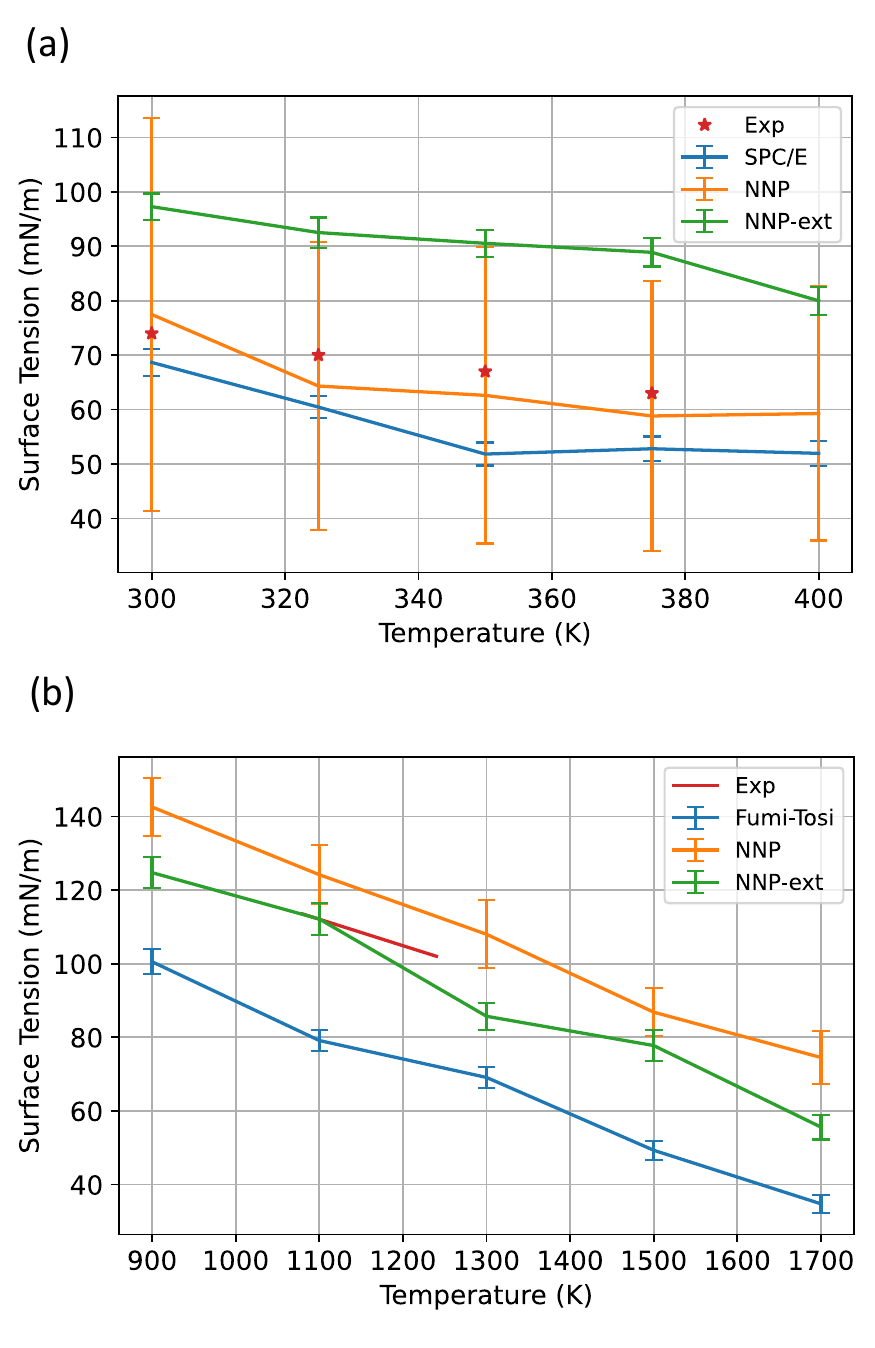}
\caption{
Average surface tension with error bars as a function of temperature for (a) water in a temperature range of 300 K to 400 K, and (b) molten NaCl in a range of 900 K to 1700 K, using classical, NNP and NNP-ext and compared to the experiments\cite{janzMoltenSaltsData1980,speightLangeHandbookChemistry2005}.}
\label{fig:nacl_gamma-T}
\end{figure}

One of the simplest inhomogeneous configurations that can be directly compared to experiment, is the liquid-vapor interface. 
The surface tension, which can be defined as the free energy per unit area of forming a liquid-vapor interface, can serve as a quantity used to assess the quality of the energetics learned by the NNP and NNP-ext respectively. 

To calculate surface tension, we run NVT MD simulations with a planar interface set up between the liquid and vacuum (low density vapor) for 3 ns, and disregard the fist half of the simulation where system is not fully in the equilibrium.
For an interface with a normal vector along the $z$ axis, the surface tension is given by
\begin{equation}
\gamma = \frac{L_{z}}{2}\langle P_{N} -P_{T}\rangle
\end{equation}
where $L_{z}$ is the $z$ dimension of the simulation box, $P_{N} = P_{zz}$ is the normal component of the pressure tensor and $P_{T} = (P_{xx}+P_{yy})/2$ is its tangential component. 

Surface tension as a function of temperature for water and molten NaCl are displated in Fig.~\ref{fig:nacl_gamma-T}a and \ref{fig:nacl_gamma-T}b, respectively. 
Similar to the trend we observed for the relative force uncertainty, the difference between the NNP and NNP-ext is drastic for water, but mostly negligible for molten NaCl. 

For water, the error bars for the NNP can be as large as 70 mN/m, about the same as the magnitude of the average of NNP values. 
Such high error bars make the NNP estimation highly unreliable. However, the NNP-ext has precise estimations for all temperatures with error bars of 2-2.5 mN/m, and shows a clear monotonic trend consistent with experiment. 
In particular, going from 300 K to 380 K, the surface tension decreases by 9 mN/m, which is about the same as the experimentally reported decline of 8 mN/m \cite{speightLangeHandbookChemistry2005}. 
Although NNP-ext performs better than NNP, it clearly overestimates the surface tension at all temperatures. 
Besides the choice of DFT parameters, this deviation from experiment can be related to the fact that in the training procedure of NNP-ext, all inhomogeneous samples are taken from externally attractive/repulsive potentials. 
The behavior of water molecules at the liquid-vapor interface is different than its corresponding response to the attractive/repulsive potentials. 
For molten NaCl, we note that while the regular NNP has larger errors bars than NNP-ext, both potentials appear to be closer to experiment than the classical Fumi-Tosi model.

\subsection{Cavitation free energies}

Finally, we consider inhomogeneous fluids in configurations most relevant to solvation, where the fluid is excluded from a cavity occupied by the solute.
We can calculate the free energy $\Delta G_{cav}$ to form a spherical cavity of radius $R$ within the liquid.
This cavitation free energy initially scales with the volume of the cavity ($\propto R^{3}$) for small $R$ and with the surface area of the cavity ($\propto R^{2}$) at larger $R$.
In particular, in the $R \rightarrow \infty$ limit, the cavitation free energy should approach $4\pi R^2 \gamma$, where $\gamma$ is the surface tension.\cite{huangScalingHydrophobicSolvation2001}
The entire profile of $\Delta G_{cav}$($R$) can provide more detail about the inhomogeneous response of fluids relevant for solvation than the surface tension value. 
Consequently, we calculate and compare this quantity for each of the three NNPs.

The cavitation free energy can be calculated through umbrella sampling \cite{kastnerUmbrellaSampling2011} and the multiple histogram method,\cite{kumarWeightedHistogramAnalysis1992} where a purely repulsive potential is used to bias the simulation towards forming larger and larger cavities.\cite{huangScalingHydrophobicSolvation2001,sundararamanRecipeFreeenergyFunctionals2014}. 
To do this, we:
\begin{enumerate}
\item initiate and equilibrate a bulk liquid configuration,
\item run a sequence of NPT calculations with spherical repulsive potentials applied at increasing radii, and
\item collect energies and values of cavity radius $\xi$ (which effectively serves as the reaction coordinate/collective variable for this free energy method), which allows us to generate histograms relevant for computing the free energy profile
\end{enumerate}

We combine the umbrella sampling and multiple histogram methods to determine  $\Delta G_{cav}(R)$ per unit surface area.(See Ref.~\citenum{huangScalingHydrophobicSolvation2001} for a detailed discussion of the procedure.)

\begin{figure}
\centering
\includegraphics[width=\columnwidth]{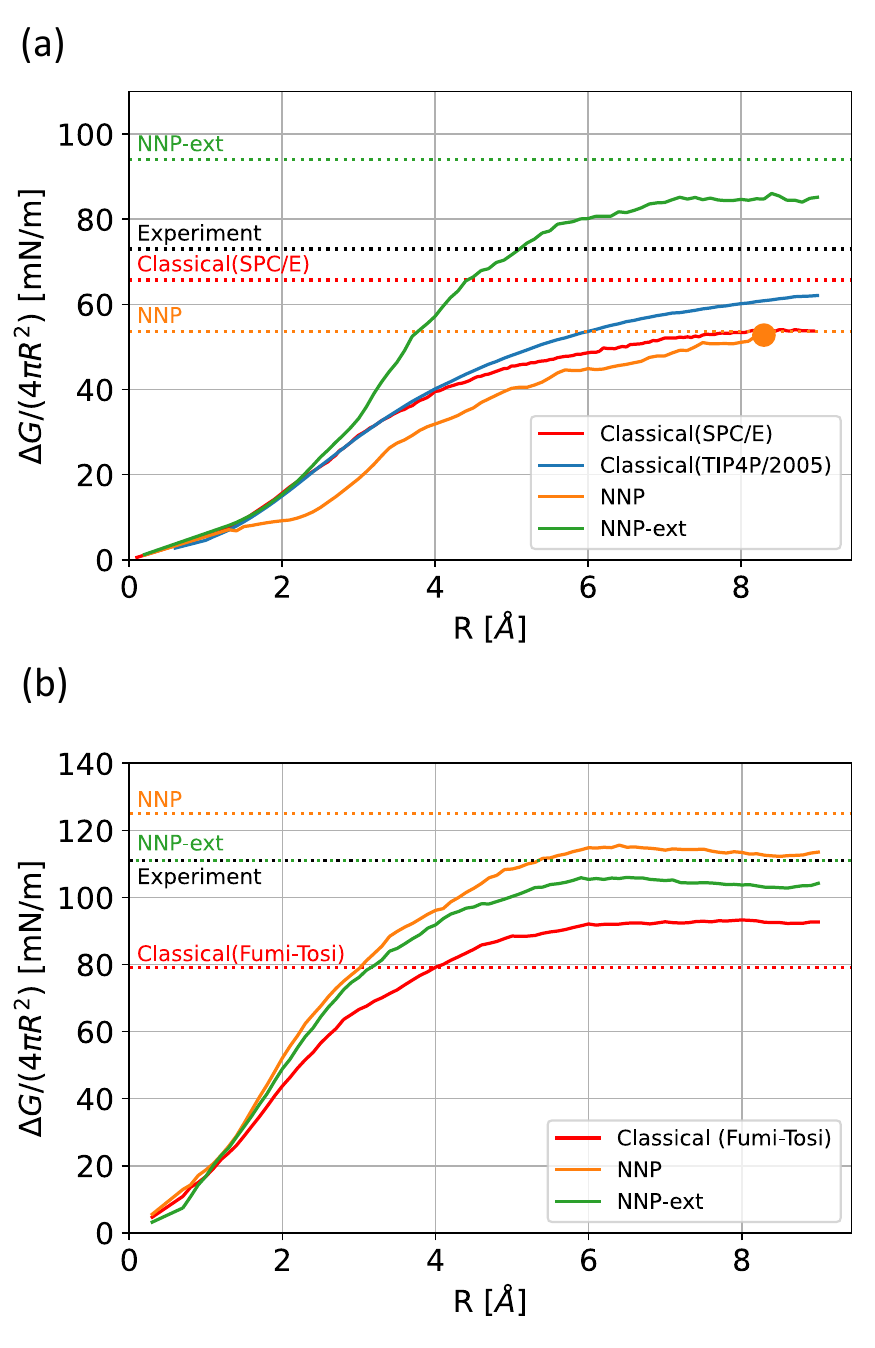}
\caption{Cavitation free energy per surface area for (a) water at 300 K and 1 bar and (b) molten NaCl at 1100 K and 1 bar, as estimated by umbrella sampling with the multiple histogram method.
The behaviour transitions from a volume proportional free energy for small radii to a constant surface tension expected at macroscopic dimensions.
Dashed lines represent the planar surface tensions, which are expected to be the asymptotic result as $R\to\infty$.}
\label{fig:GibbsCavity}
\end{figure}

From these simulations, we compute $\Delta G_{cav}$,
\begin{equation}
G=-kT\cdot ln\left(Q\right)
\end{equation}
with the partition function,
\begin{equation}
Q\left(\xi\right)=\frac{\int\delta\left[\xi\left(r\right)-\xi\right]exp\left[-\frac{E\left(r\right)}{kT}\right]dr}{\int exp\left[-\frac{E\left(r\right)}{kT}\right]dr}.
\end{equation}

We obtain a mean estimate of $\Delta G_{cav}$ and its variance from each simulation, and then use the multiple histogram method to combine all the estimates into a single $\Delta G_{cav}(R)$ curve.

Figures~\ref{fig:GibbsCavity} a) and b) respectively display our calculated $\Delta G_{cav}(R)$ profiles for water at the top and molten NaCl at the bottom, for different types of interatomic potential. 
In agreement with previous results for molecular liquids,\cite{sundararamanRecipeFreeenergyFunctionals2014} at small radii the cavitation free energy is proportional to the volume of the cavity $N_{bulk}T\times4\pi R^{3}/3$, because the volume dependence at small scales is essentially the probability of finding an empty sphere within the liquid. 
Note that behavior in this regime is independent of the interatomic potential.

As the cavity radius increases, the details of the interatomic potential become more important and lead to deviations between the different cases. 
At a large enough radius, the interface between the cavity and liquid is approximately planar, and the result should approach the surface tension of the liquid/vapor interface. 

For water, the classical SPC/E and TIP4P underestimate the free cavitation energy, in accordance with previous studies.\cite{sundararamanComputationallyEfficaciousFreeenergy2012,sundararamanRecipeFreeenergyFunctionals2014}
TIP4P/2005 performs better at larger radii and reach 62 mN/m at 9 \AA. 
The regular NNP predicts even lower cavitation energies for radii larger than 1.7 \AA, and fails to complete trajectories beyond radii of 8.2 \AA. 
In contrast, the NNP-ext method including inhomogeneous configurations in the training data is much more stable and slightly overestimates the asymptotic surface tension (84.5 mN/m) compared to the experimental values (74 mN/m). 

The overall trend of the cavitation free energy is comparable to the previous studies with classical MD and classical DFT, and the slight overestimation is anticipated due to the overstructuring of water in the underlying AIMD. It should be mentioned that this overestimation is still in a better agreement to the experimental value than the liquid-vapor interface estimation from the previous section (dotted line).

For molten NaCl, all three methods produce overall reasonable profiles, but the classical MD underestimates the overall magnitude, while both NNPs reach an asymptote in much better agreement with experiment as discussed above.

\section{Conclusion}

We have developed techniques to train neural network potentials that are better suited for the modeling of inhomogeneous liquids.
Specifically, we use external potentials to bias the liquid configurations to introduce different classes of inhomogeneities, and use this approach both in generating the training data and in testing the resulting potentials.
We test NNPs and classical potentials in their density response to external potentials, surface tension and cavitation free energy for two different types of liquids, the molecular liquid water, and the simple ionic liquid molten NaCl. For water, our results show that NNPs trained to bulk liquids alone underperforms, and sometimes even fail to yield a stable trajectory.
In contrast, NNP-exts trained to AIMD data including inhomogeneous configurations can substantially improve the predictions of cavitation free energies and surface tensions compared to classical potentials. However, for the molten salt, both NNP and NNP-ext predict almost identical properties, with a slightly better predictions of NNP-ext for surface tension. This contrasting response of water and molten NaCl can be explained by the complexity of the liquids. Water with intramolecular covalent bonds, hydrogen bonds and long range interactions cannot be modelled with only homogeneous samples and a diverse set of inhomogeneous samples is necessary to train a reliable NNP. On the the other hand, for simple liquids such as molten NaCl, homogeneous samples are enough to capture the ionic interactions in inhomogeneous environments. 

This study provide a pathway to systematically map the response of inhomogeneous fluids and inform the construction of classical density functionals based on long-equilibrated MD trajectories with NNPs.

\section*{Acknowledgements}

This work was supported by the U.S. Department of Energy, Office of Science, Basic Energy Sciences, under Award No. DE-SC0022247.
Calculations were carried out at the Center for Computational Innovations at Rensselaer Polytechnic Institute, and at the National Energy Research Scientific Computing Center (NERSC), a U.S. Department of Energy Office of Science User Facility located at Lawrence Berkeley National Laboratory, operated under Contract No. DE-AC02-05CH11231 using NERSC award ERCAP0020105.

\appendix
\section{Interatomic potential details}

\subsection{Classical potentials}\label{sec:ClassicalPot}

We use the 3-site extended simple point charge (SPC/E) for water which consists of a Lennard-Jones sphere at the oxygen atom and 3 point charges at each of the atomic sites.\\

We use the standard additive pairwise Fumi-Tosi rigid ion model for NaCl  with pairwise interactions based on the parameters in reference \cite{fumiIonicSizesBorn1964} .

\subsection{Neural network potentials}\label{sec:NNP}

To train the NNPs, we use DeePMD-kit (version 2.0.3) \cite{wangDeePMDkitDeepLearning2018}with the training data generated from AIMD simulations through JDFTx \cite{sundararamanJDFTxSoftwareJoint2017}. 

We select the Deep Potential Smooth Edition descriptor with two atom embeddings, radial cut-off was 6 \AA, 16 axis neurons, and embedding NN of shape [25, 50, 100].
The fitting of the position descriptor is fed into an NN of shape [240, 240, 240] and trained to energy and force, weighted 0.02:1000 respectively as initial loss prefactors.
The initial exponential decay learning rate is 0.001.
In general, the NN hyperparameters are sampled in layers and neurons to achieve acceptable errors. 

The AIMD training data should include configurations at state points to be interpolated when using the NNP in MD. 
The configurations in the training set are snapshots of atomic positions and resulting forces on each atom.

For the water NNP, the training data includes water configurations of 192 atoms run for 2~ps in AIMD at different temperatures and pressures shown in Table \ref{tab:water-training-sets}. 
In this paper, we choose the SCAN  functional based on its accuracy in previous studies\cite{zhengStructuralElectronicDynamical2018}.
Training errors at the end of training are 0.001 eV RMSE for energy and 0.1 eV/\AA for forces.

\begin{table} \centering
\caption{Structure, thermodynamic state points, and number of configurations sampled (10~fs apart) for each water AIMD simulation used to train the NNPs utilized in this study.}
\label{tab:water-training-sets}
\begin{tabular}{lccc}
\hline\hline
Structure  & $T$ (K)  & $P$ (bar)  & \# configs \\
\hline
Liquid & 300 & 1 & 201 \\
Liquid  & 300 & 1000 & 201 \\
Liquid  & 330 & 1 & 201 \\
Liquid & 400 & 10 & 201 \\
Liquid & 400 & 1000 & 201 \\
Vapour & 500 & 1 & 201 \\
Liquid  & 500 & $10^{3}$ & 201 \\
\hline\hline
\end{tabular}
\end{table}

For the NaCl NNP, the training data includes NaCl configurations of 64 atoms run for 2~ps in AIMD at different temperatures as solid, liquid, and at high pressure. 
See Table II in reference \cite{shahFirstprinciplesMoltenSaltTBD} for the detailed list of simulations considered in training and AIMD parameters used. 
In this paper, we choose the PBE-D2 \cite{grimmeSemiempiricalGGAtypeDensity2006} functional based on its past accuracy. \cite{shahFirstprinciplesMoltenSaltTBD}
Training errors at the end of training are 0.006 eV RMSE for energy and 0.03 eV/\AA for forces.

\begin{table} \centering
\caption{Structure, thermodynamic state points, number of configurations sampled (10~fs apart), and external planar Gaussian potential for each water AIMD simulation.}
\label{tab:waterPert-training-sets}
\begin{tabular}{lcccc}
\hline\hline
Structure  & $T$ (K)  & $P$ (bar) & $\pm U$ (eV)  & \# configs \\
\hline
Liquid & 330 & 1 & 1.36 & 201 \\
Liquid & 330 & 1 & 2.72 & 201 \\
Liquid & 330 & 1 & 5.44 & 201 \\
Liquid & 400 & 10 & 1.36 & 201 \\
Liquid & 400 & 10 & 2.72 & 201 \\
Liquid & 400 & 10 & 5.44 & 201 \\
Liquid & 500 & 100 & 1.36 & 201 \\
Liquid & 500 & 100 & 2.72 & 201 \\
Liquid & 500 & 100 & 5.44 & 201 \\
\hline\hline
\end{tabular}
\end{table}

The training data for the NNP-ext for water includes the standard NNP water data plus a range of repulsive and attractive planar Gaussian potentials varying along the Z axis. 
The additional external potential training data is shown in Table \ref{tab:waterPert-training-sets}.
Training errors at the end of training are 0.007 eV RMSE for force and 0.1 eV/\AA for energy.
Similarly, the training data for the NNP-ext for NaCl includes the standard NaCl data plus a range of repulsive and attractive planar Gaussian potentials varying along the Z axis applied to Na. 
The additional external potential training data is shown in Table \ref{tab:NaClPert-training-sets}.
Training errors at the end of training are 0.01 eV RMSE for force and 0.03 eV/\AA for energy.

\begin{table} \centering
\caption{Structure, thermodynamic state points, number of configurations sampled (10~fs apart), and external planar Gaussian potential (applied to Na) for each NaCl AIMD simulation.}
\label{tab:NaClPert-training-sets}
\begin{tabular}{lcccc}
\hline\hline
Structure  & $T$ (K)  & $P$ (bar) & $\pm U$ (eV)  & \# configs \\
\hline
Liquid & 1500 & 1 & 0.544 & 201 \\
Liquid & 1500 & 1 & 0.816 & 201 \\
Liquid & 1500 & 1 & 1.360 & 201 \\
Liquid & 1500 & 1 & 3.264 & 201 \\
Liquid & 1500 & 1 & 6.800 & 201 \\
\hline\hline
\end{tabular}
\end{table}

\section*{References}
\bibliographystyle{IEEEtran}

\end{document}